\documentclass[prb,twocolumn]{revtex4}%

\usepackage{graphicx}
\usepackage{amsmath}
\usepackage{ulem}

\newcommand{\be}{\begin{equation}}
\newcommand{\ee}{\end{equation}}
\newcommand{\bea}{\begin{eqnarray*}}
\newcommand{\eea}{\end{eqnarray*}}
\newcommand{\bean}{\begin{eqnarray}}
\newcommand{\eean}{\end{eqnarray}}

\begin{document}

\draft
\title
{\bf Current Rectification and Seebeck Coefficient of Serially
Coupled Double Quantum Dots}

\author{ Yen-Chun Tseng and David M.-T. Kuo$^{*}$ }
\address{Department of Electrical Engineering, National Central
University, Chungli 320 Taiwan}


\begin{abstract}
The transport properties of serially coupled quantum dots (SCQDs)
embedded in a matrix connected to metallic electrodes are
theoretically studied in the linear and nonlinear regimes. The
current rectification and negative differential conductance of SCQDs
under the Pauli spin blockade condition are attributed to the
combination of bias-direction dependent probability weight and
off-resonant energy levels yielded by the applied bias across the
junctions. We observe the spin-polarization current rectification
under the Zeeman effect. The maximum spin-polarization current
occurs in the forward bias regime. Such behavior is different from
the charge current rectification. Finally, the Seebeck coefficient
($S$)of SCQDs is calculated and analyzed in the cases without and
with electron phonon interactions. The application of SCQDs as a
temperature detector is discussed on the basis of the nonlinear
behavior of $S$ with respect to temperature difference across the
junction.
\end{abstract}


\maketitle


\textbf{1. Introduction}

Serially coupled quantum dots (SCQDs) exhibit the transport
properties of current rectification due to the Pauli spin blockade,
negative differential conductance (NDC), nonthermal broadening of
the tunneling current, and coherent tunneling in the Coulomb
blockade regime.$^{1-3)}$ Although many theoretical works have been
devoted to investigating these phenomena, they still can not explain
the transport properties of SCQDs systematically.$^{4-8)}$ Sun e al
calculated the tunneling current of SCQDs in the Pauli spin blockade
using the Keldysh-Green function technique.$^{4)}$ The procedure
introduced in ref [4] to solve high-order Green functions arising
from electron Coulomb interactions can not resolve the quantum paths
of SCQDs. In Refs. 5-8, the master equation was used to calculate
the tunneling current of SCQDs. However, in these works, cases are
restricted to $t_c \ll \Gamma$, where $t_c$ and $\Gamma$ denote,
respectively, the interdot hopping strength and tunneling rate
between the electrodes and the quantum dots (QDs).

Here, a closed-form expression for the tunneling current of SCQDs
with a finite interdot hopping strength enables the analysis of
current rectification arising from coherent tunneling with a spin
blockade and the NDC of tunneling current resulting from
off-resonant energy levels. The effect of Zeeman energy splitting on
tunneling current is investigated to clarify the behavior of
spin-polarization current. In addition, the Seebeck coefficient
($S$) of SCQDs is calculated in the cases without and with electron
phonon interactions (EPIs). In the absence of EPIs, we propose how
to use SCQDs as a temperature detector on the basis of the nonlinear
Seebeck coefficient. When the SCQDs are embedded in a phonon
cavity$^{9-11)}$, it is possible to manipulate EPIs to control the
electrical conductance and Seebeck coefficient of junction systems.


\textbf{2. Formalism}

Because we consider nanoscale semiconductor QDs, the energy level
separation of between QDs is much larger than their on-site Coulomb
interactions and thermal energies. One energy level for each quantum
dot is considered in this study. The two-level Anderson model
including EPIs is employed to simulate the SCQD junction system
shown in the inset of Fig. 1(a). The Hamiltonian of an SCQD
junction$^{12)}$ is given by $H=H_{0}+H_{CQD}+H_T$

\begin{eqnarray}
H_0& = &\sum_{k,\sigma} \epsilon_k
a^{\dagger}_{k,\sigma}a_{k,\sigma}+ \sum_{k,\sigma} \epsilon_k
b^{\dagger}_{k,\sigma}b_{k,\sigma}\\ \nonumber &+&\sum_{k,\sigma}
V_{k,1}d^{\dagger}_{1,\sigma}a_{k,\sigma}
+\sum_{k,\sigma}V_{k,2}d^{\dagger}_{2,\sigma}b_{k,\sigma}+c.c
\end{eqnarray}
where the first two terms describe the free electron gas of the left
and right metallic electrodes. $a^{\dagger}_{k,\sigma}$
($b^{\dagger}_{k,\sigma}$) creates  an electron with momentum $k$
and spin $\sigma$ with energy $\epsilon_k$ in the left (right)
metallic electrode. $V_{k,\ell}$ ($\ell=1,2$) describes the coupling
between the metallic electrodes and the first (second) QD.
$d^{\dagger}_{\ell,\sigma}$ ($d_{\ell,\sigma}$) creates (destroys)
an electron in the $\ell$-th dot.

\begin{small}
\begin{eqnarray}
H_{CQD}&=& \sum_{\ell,\sigma} E_{\ell} n_{\ell,\sigma}+
\sum_{\ell} U_{\ell} n_{\ell,\sigma} n_{\ell,\bar\sigma}\\
\nonumber &+&\frac{1}{2}\sum_{\ell,j,\sigma,\sigma'}
U_{\ell,j}n_{\ell,\sigma}n_{j,\sigma'}
+\sum_{\ell,j,\sigma}t_{\ell,j} d^{\dagger}_{\ell,\sigma}
d_{j,\sigma},
\end{eqnarray}
\end{small}
where $E_{\ell}$ is the spin-independent QD energy level and
$n_{\ell,\sigma}=d^{\dagger}_{\ell,\sigma}d_{\ell,\sigma}$.
Notations $U_{\ell}$ and $U_{\ell,j}$ describe the intradot and
interdot Coulomb interactions, respectively. $t_{\ell,j}$ describes
the electron interdot hopping. $H_T$ describes the EPIs

\begin{equation}
H_T= \omega_0 c^{\dagger} c+ \sum_{\ell,\sigma} \Omega_{\ell}
n_{\ell,\sigma} (c^{\dagger}+c),
\end{equation}
where $\omega_0$ is the phonon cavity frequency and $\Omega_{\ell}$
is the coupling strength of EPIs. A canonical transformation can be
carried out to remove on-site EPIs, that is, $H_{new}=
e^{S^{\dagger}} H e^S$, where $S=-\sum_{\ell,\sigma}\Omega_{\ell}
n_{\ell,\sigma} (c^{\dagger}-c)$.$^{13,14)}$ In the new Hamiltonian,
we have the following effective physical parameters:
$V^e_{k,1}=V_{k,1}e^{\lambda_1(c^{\dagger}-c)}$,
$V^e_{k,2}=V_{k,2}e^{\lambda_2 (c^{\dagger}-c)}$,
$E^e_{\ell}=E_{\ell}-\lambda^2_{\ell}\omega_0$,
$U^e_{\ell}=U_{\ell}-2\lambda^2_{\ell} \omega_0$,
$U^e_{\ell,j}=U_{\ell,j}-2\lambda_1 \lambda_2 \omega_0$,
$t^e_{\ell,j}=t_{\ell,j}
e^{-(\lambda_{\ell}-\lambda_j)(c^{\dagger}-c)}$, and
$\lambda_{\ell}=\Omega_{\ell}/\omega_0$. Under the canonical
transformation, the coupling strengths between the electrodes and
the dots, on-site energy levels, intradot Coulomb interactions,
interdot Coulomb interactions, and electron interdot hopping
strengths are renormalized by EPIs. If we consider a special case of
$\lambda_1=-\lambda_2=\lambda$, we have
\begin{equation}
H_T= \omega_0 c^{\dagger} c+ \sum_{\sigma} \Omega
(n_{1,\sigma}-n_{2,\sigma}) (c^{\dagger}+c).
\end{equation}
This special case of Eq. (4) was already considered in refs 15 and
16. For the case of Eq. (4), we have an effective electron interdot
Coulomb interaction $U_{\ell,j}+2\lambda^2 \omega_0$, which is
always repulsive and enhanced with increasing EPIs.$^{16)}$

To decouple the EPIs of $H_{new}$, we take the mean-field average to
remove the phonon field arising from $c^{\dagger}-c$, which is $<
exp^{\lambda_{\ell}(c^{\dagger}-c)} >
=exp^{-\frac{1}{2}\lambda^2_{\ell} coth^2[\omega_0/(2k_BT_p)]}$. On
the basis of such a mean-field average, we see a reduction of
$V_{k,\ell}exp^{-\frac{1}{2}\lambda^2_{\ell}
coth^2[\omega_0/(2k_BT_p)]}=V_{k,\ell}X_{\ell}$ and interdot hopping
strength $t_c exp^{-\frac{1}{2}(\lambda_{1}-\lambda_2)^2
coth^2[\omega_0/(2k_BT_p)]}=t_cX_{\ell,j}$. This leads to the
redefinition of the renormalized resonance width of each energy
level of the QD. $T_p$ is the phonon temperature.

Using the Keldysh-Green function technique and neglecting the phonon
-assisted tunneling process,$^{12)}$ the tunneling current of an
SCQD is given by
\begin{equation}
J=\frac{2e}{h}\int d\epsilon {\cal T}(\epsilon)
[f_L(\epsilon)-f_R(\epsilon)],
\end{equation}
where ${\cal T}(\epsilon)\equiv ({\cal T}_{12}(\epsilon) +{\cal
T}_{21}(\epsilon))/2$ is the transmission factor.
$f_{L=1(R=2)}(\epsilon)=1/[e^{(\epsilon-\mu_{L(R)})/k_BT_{L(R)}}+1]$
denotes the Fermi distribution function for the left (right)
electrode. The left (right) chemical potential is given by
$\mu_L(\mu_R)$. $\mu_L-\mu_R=e\Delta V_a$, where $\Delta V_a$
denotes the applied bias. Notation $T_{L(R)}$ denotes the
equilibrium temperature of the left (right) electrode. $e$ and $h$
denote the electron charge and Planck's constant, respectively.
${\cal T}_{\ell,j}(\epsilon)$ denotes the transmission coefficient,
which can be calculated by the on-site retarded Green function and
the lesser Green function. The transmission coefficient has the
following expression:
\begin{equation}
{\cal
T}_{\ell,j}(\epsilon)=-2\sum^{8}_{m=1}\frac{\Gamma^e_{\ell}(\epsilon)
\Gamma^{e,m}_{j}(\epsilon)}{\Gamma^e_{\ell}(\epsilon)+\Gamma^{e,m}_{j}(\epsilon)}
\mbox{Im}G^r_{\ell,m,\sigma}(\epsilon),
\end{equation}
where Im means taking the imaginary part of the function that
follows, and
\begin{equation}
G^r_{\ell,m,\sigma}(\epsilon)=p_m/(\mu_{\ell}-\Sigma_m).
\end{equation}
$\Gamma^e_{ \ell=L(1),R(2)}(\epsilon)=\Gamma_{\ell}X_{\ell}^2$,
where $\Gamma_{\ell}=\sum_{k}V^2_{k,\ell} \delta
(\epsilon-\epsilon_k)$ denotes the tunnel rate from the left
electrode to dot A ($E_1$) and from the right electrode to dot B
($E_2$), which is assumed to be energy- and bias-independent for
simplicity. $\mu_{\ell}=\epsilon-E^e_{\ell}+i\Gamma^e_{\ell}/2$. We
can assign the following physical meaning to Eq. (6). The sum in Eq.
(6) is over eight possible configurations labeled by $m$. We
consider an electron (of spin $\sigma$) entering level $\ell$, which
can be either occupied (with probability $N_{\ell,\bar\sigma}$) or
empty (with probability $1-N_{\ell,\bar\sigma}$). For each case, the
electron can hop to level $j$, which can be empty (with probability
$a_j=1-N_{j,\sigma}-N_{j,\bar\sigma}+c_j$), singly occupied in a
spin $\bar\sigma$ state (with probability
$b_{j,\bar\sigma}=N_{j,\bar\sigma}-c_j$) or spin $\sigma$ state
(with probability $b_{j,\sigma}=N_{j,\sigma}-c_j$), or in a
double-occupied state (with probability $c_j$). Thus, the
probability factors associated with the eight configurations
appearing in Eq. (6) are $p_1=(1-N_{\ell,\bar\sigma})a_j$,
$p_2=(1-N_{\ell,\bar\sigma})b_{j,\bar\sigma}$,
$p_3=(1-N_{\ell,\bar\sigma})b_{j,\sigma}$,
$p_4=(1-N_{\ell,\bar\sigma})c_j$, $p_5=N_{\ell,\bar\sigma}a_j$,
$p_6=N_{\ell,\bar\sigma}b_{j,\bar\sigma}$,
$p_7=N_{\ell,\bar\sigma}b_{j,\sigma}$, and
$p_8=N_{\ell,\bar\sigma}c_j$. $\Sigma_m$ in the denominator of Eq.
(7) denotes the self-energy correction due to Coulomb interactions
and coupling with level $j$ (which couples with the other electrode)
in configuration $m$. We have $\Sigma_1=t^{2}/\mu_j$,
$\Sigma_2=U^e_{\ell,j}+t^{2}/(\mu_j-U^e_j)$,
$\Sigma_3=U^e_{\ell,j}+t^{2}/(\mu_j-U^e_{j,\ell})$,
$\Sigma_4=2U^e_{\ell,j}+t^{2}/(\mu_j-U^e_j-U^e_{j,\ell})$,
$\Sigma_5=U^e_{\ell}+t^{2}/(\mu_j-U^e_{j,\ell})$,
$\Sigma_6=U^e_{\ell}+U^e_{\ell,j}+t^{2}/(\mu_j-U^e_j-U^e_{j,\ell})$,
$\Sigma_7=U^e_{\ell}+U^e_{\ell,j}+t^{2}/(\mu_j-2U^e_{j,\ell})$, and
$\Sigma_8=U^e_{\ell}+2U^e_{\ell,j}+t^{2}/(\mu_j-U^e_j-2U^e_{j,\ell})$.
Note that $t=t^e_c=t_cX_{\ell,j}$. Here $\Gamma^{e,m}_{j}=-2$Im$
\Sigma_j$ denotes the effective tunneling rate from level $l$ to the
other electrode through level $j$ in configuration $m$. For example,
$\Gamma^{e,1}_{j}=-2$Im$t^{2}/\mu_j=t^{2}\Gamma^e_j/[(\epsilon-E^e_j)^2+(\Gamma^e_j/2)^2]$.
It is noted that $\Gamma^{e,m}_{j}$ has the numerator $\Gamma^e_j$
for all configurations. Furthermore,
$G^r_{\ell,\sigma}(\epsilon)=\sum^{8}_{m=1}
G^r_{\ell,m,\sigma}(\epsilon)$ is simply the on-site single-particle
retarded Green function for level ${\ell}$ as given in Eq. (A16) in
Ref. 12, and $G^r_{\ell,m,\sigma}(\epsilon)$ corresponds to its
partial Green function in configuration $m$.

The probability factors of Eq. (7) are determined by the thermally
averaged one-particle occupation number and two-particle correlation
functions, which can be obtained by solving the on-site lesser Green
functions:$^{12)}$

\begin{equation}
N_{\ell,\sigma}=-\int \frac{d\epsilon}{\pi}\sum^8_{m=1}
\frac{\Gamma^e_{\ell}f_{\ell}(\epsilon)+\Gamma^{e,m}_jf_j(\epsilon)}{\Gamma^e_{\ell}+\Gamma^{e,m}_{j}}
ImG^r_{\ell,m,\sigma}(\epsilon),
\end{equation}

and
\begin{equation}c_{\ell}=-\int
\frac{d\epsilon}{\pi}\sum^8_{m=5}
\frac{\Gamma^e_{\ell}f_{\ell}(\epsilon)+\Gamma^{e,m}_jf_j(\epsilon)}{\Gamma^e_{\ell}+\Gamma^{e,m}_{j}}
ImG^r_{\ell,m,\sigma}(\epsilon).
\end{equation}
Note that $\ell \neq j$ in Eqs. (6), (8) and (9), which are valid
under the condition of $t_c/U_{\ell} \ll 1$. We will study the
transport properties of SCQDs on the basis of Eqs. (5), (8), and
(9).


\textbf{3. Results and discussion}

\textbf{3.1 Current rectification}

To numerically calculate the tunneling current of SCQDs without EPIs
($\lambda_1=\lambda_2=0$), we adopt the intradot Coulomb
interactions of $U_{\ell}=U$ and tunneling rates of
$\Gamma_L=\Gamma_R=\Gamma$. We consider these conditions of
homogenous intradot electron Coulomb interactions, and symmetrical
tunneling rates for simplicity. All energy scales are considered in
units of $\Gamma_0$. On the other hand, $\eta_{1(2)}e\Delta V_a$ is
employed to describe the energy shift arising from the applied bias
$\Delta V_a$ across the junction. That means  that $E_{\ell}$ is
replaced by $\epsilon_{\ell}=E_{\ell}+\eta_{\ell} e\Delta V_a$,
assuming the right electrode is grounded. On the basis of the
experiment in Ref. 1, we adopt $\eta_1=0.6$ and $\eta_2=0.4$.
Although the factor $\eta_{\ell}$ depends on the QD shape, material
dielectric constant, and location, we assume that $\eta_{\ell}$ is
determined by the QD location, that is $\eta_{\ell}= L_{\ell}/L$,
where $L_{\ell}$ is the distance between the grounded electrode and
the $\ell$th QD, and $L$ is the separation distance between the left
electrode and the right electrode.

We plot the tunneling current of SCQDs under the Pauli spin blockade
condition ($E_1+U_{12}=E_2+U_2$) shown in the inset of Fig. 1(a) for
three different interdot Coulomb interactions at $k_BT=1~\Gamma_0$,
$t_c=0.1~\Gamma_0$, $U=30~\Gamma_0$, and $\Gamma=~\Gamma_0$. The
three curves correspond to (a) $U_{12}=0$ and $E_1=E_F$, (b)
$U_{12}=5~\Gamma_0$ and $E_1=E_F-5~\Gamma_0$, and (c)
$U_{12}=10~\Gamma_0 $ and $ E_1=E_F-10~\Gamma_0$. $E_F$ denotes the
Fermi energy of the electrodes. The current rectification and
negative differential conductance (NDC) of SCQDs are observed. The
maximum tunneling current is suppressed in the presence of interdot
Coulomb interactions. We note that the $J_{max,R}/J_{max,F}$ ratios
are near 2. $J_{max,R}$ and $J_{max,F}$ are the maximum current in
the backward bias and forward bias, respectively.
$J_{max,R}/J_{max,F}=2$ is in very good agreement with an
experimental observation.$^{1}$ The results shown in Fig. 1 indicate
that the Pauli spin blockade condition is not attributed to interdot
Coulomb interactions, but to intradot Coulomb interactions. On the
basis of Eq. (6), the Pauli spin blockade resonant channel of
$E_1+U_{12}=E_2+U_2$ is determined from the probability weights of
$p_2=(1-N_{1,\bar\sigma})(N_{2,\bar\sigma}-c_2)$ and
$p_5=N_{2,\bar\sigma}(1-N_{1,\bar\sigma}-N_{1,\sigma}+c_1)$, which
result from ${\cal T}_{12}$ and ${\cal T}_{21}$, respectively.
Figure 1(b) shows the occupation number
$N_{1(2),\bar\sigma}=N_{1(2),\sigma}=N_{1(2)}$ and two particle
correlation functions $c_{1(2)}$ for $U_{12}=10\Gamma_0$. The $p_2$
of ${\cal T}_{12}$ and $p_5$ of ${\cal T}_{21}$ are also plotted as
indicated by the dotted line and dashed line in Fig. 1(b),
respectively. Because $J_{max,F}$ and $J_{max,R}$ occur,
respectively, at $e\Delta V_a=3\Gamma_0$ and $e\Delta
V_a=-4.5\Gamma_0$, we have $p_2=0.28$ and $p_5=0.3$ at $e\Delta
V_a=3\Gamma_0$ and $p_2=0.24$ and $p_5=0.65$ at $e\Delta
V_a=-4.5\Gamma_0$. Their sum is $p_2+p_5=0.58$ at $J_{max,F}$ and
$p_2+p_5=0.89$ at $J_{max,R}$. This demonstrates why $J_{max,R}$ is
larger than $J_{max,F}$. In the reversed bias, the increase in $p_5$
with respect to $e\Delta V_a$ results from the enhancement of
$(1-N_{1,\bar\sigma}-N_{1,\sigma}+c_1)$ (dot A empty), which
provides an increased probability of tunneling of electrons in the
right electrode. This result of $p_5$ approaching one in the high
backward bias indicates that the current spectra shown in Fig. 1(a)
are determined not only by the probability weights of $p_2$ and
$p_5$ but also by off-resonant energy levels. We find that the
off-resonant energy level is a key reason to observe the NDC
behavior of SCQDs. As a consequence, the QD energy level broadening
will significantly influence the maximum currents in the forward and
reversed biases.$^{3)}$ It is worth noting that the phonon-assisted
tunneling process arising from EPIs can not be ignored in the high
bias regime.$^{7)}$ Some current structures of SCQDs$^{1)}$ in the
high-bias regime can be well explained by the phonon-assisted
tunneling process.$^{7)}$

For the applications of SCQDs in spintronics, it is crucial to
measure the spin configuration of each electron in individual QDs.
SCQDs have a functionality of spin-charge conversion.$^{1-3)}$
However, it is not easy to measure the small magnitude of tunneling
current in the weak interdot coupling limit of $t_c/\Gamma \ll 1$.
Although tunneling current can be enhanced by increasing $t_c$, we
consider how the behavior of current rectification is influenced for
$t_c/\Gamma \approx 1 $. To clarify the above question, we plot the
tunneling current of SCQDs at various $t_c$ values in Figs.
2(a)-2(c).Other physical parameters are the same as those for the
curve with $U_{12}=10\Gamma_0$ shown in Fig. 1(a). The maximum
currents labeled $J_{max,F}$ and $J_{max,R}$ are shifted toward a
higher bias when $t_c$ increases. The $J_{max,R}/J_{max,F}$ ratio
slightly decreases. The three maximum currents are $J_{max,F}$=10.4,
167.8, and 328, which correspond to $e\Delta V_a$=3, 3.6, and 4.8
$\Gamma_0$ , respectively.($J_{max,R}$=21.6, 319, and 599 correspond
to -4.5, -5.6, and -7.4$\Gamma_0$, respectively). The dashed lines
shown in Figs. 2(a)-2(c) show the contributions arising from only
the resonant channels $\Pi_2$ of ${\cal T}_{12}$ and $\Pi_5$ of
${\cal T}_{21}$. These dashed lines  very close the solid lines at
the small applied bias (between $J_{max,F}$ and $J_{max,R}$) but not
at the large applied bias. From the results shown in Figs.
2(a)-2(c), the contributions of $\Pi_2$ and $\Pi_5$ still dominate
the trend of current spectra. These two resonant channels $\Pi_2$
and $\Pi_5$ have the two poles $\epsilon_{\pm}=E_F+0.5e\Delta
V_a+i\Gamma/2 \pm 0.5 \sqrt{(0.2e\Delta V_a)^2+4t^2_c}$. Their
bias-dependent probability weights
$p_2=(1-N_{1,\bar\sigma})(N_{2,\bar\sigma}-c_2)$ and
$p_5=N_{2,\bar\sigma}(1-N_{1,\bar\sigma}-N_{1,\sigma}+c_1)$ are
determined by the occupation numbers shown in Figs. 2(d)-2(f). The
effect of interdot hopping on $N_1$ and $N_2$ is enhanced with
increasing $t_c$. The results shown in Fig. 2 indicate that the
current rectification behavior of the SCQD is not destroyed under
the condition of $t_c/\Gamma \approx 1$.

Although the SCQD system has the functionality of spin filters under
the Pauli spin blockade condition, the tunneling currents in Figs. 1
and 2 do not exhibit the spin-polarization current. The
spin-polarization current of SCQDs was theoretically studied in Ref.
4 in which the authors considered the spin-bias and ferromagnetic
electrodes. Our study introduces the Zeeman effect arising from a
local magnetic field to yield the spin-polarization current under
the Pauli spin blockade condition of $E_1+U_{12}=E_2+U_2$. Here the
energy level of each QD depends on electron spin. That is
,$E_{\ell,\sigma}=E_{\ell}\pm g_{\ell} \mu_B B \sigma=E_{\ell}\mp
\Delta_{Z,\ell,\sigma}$ is considered, where $g_{\ell}$ denotes the
$g$-factor, $\mu_B$ is the Bohr magneton, and $B$ is the magnetic
field. For simplicity, we assume the homogenous $g$ factor
$g_{\ell}=g < 0$. We plot the spin-dependent charge current
$J_{\sigma}$ ($J_{-\sigma}$) and spin-polarization current
$J_p=J_{\sigma}-J_{-\sigma}$ in Fig. 3. Other physical parameters
are the same as those for the curve of $U_{12}=10\Gamma_0$ shown in
Fig. 1. The increase in spin-polarization current is observed with
increasing Zeeman splitting. Unlike charge current, the maximum
spin-polarization current in the forward bias is larger than that in
the backward bias. That is $J_{p,max,F} > J_{p,max,B} $. Figure 3(c)
shows the spin-dependent occupation number which determines the
spin-dependent probability weights. So far, we have considered
$T_L=T_R=T$ in Figs. 1-3, where the charge currents are generated
only by the applied bias. To study the thermoelectric effect of
SCQDs, we will consider the case of $T_L \neq T_R$ to investigate
the Seebeck coefficient.

\textbf{3.2 Seebeck coefficient}

If the SCQD junction system is in an open circuit, an
electrochemical potential will form in response to a temperature
difference across junction [see Eq. (5)]; this electrochemical
potential is known as the Seebeck voltage (Seebeck effect). The
Seebeck coefficient (amount voltage generated per unit temperature
gradient ) is defined as $S=\Delta \mu/\Delta T$, where $\Delta
\mu=\mu_L-\mu_R=e\Delta V_a$ and $\Delta T=T_L-T_R$ are the voltage
difference and temperature difference across the junction,
respectively. Recently, many studies have been devoted to the
investigation of the thermoelectric effects of SCQDs in the linear
response regime.$^{17-18)}$ Because the efficiency of thermoelectric
devices can be enhanced with a large temperature difference,$^{19)}$
it is important to clarify the behavior of the Seebeck coefficient
at a large temperature difference $\Delta T$. Previous studies
focused on the nonlinear thermoelectric properties of individual QD
systems,$^{20,21)}$ which may not be readily realized from the
experimental point of view. The thermal resistivity of the SCQD
junction system can be larger than that of a single QD system. This
feature allows the SCQD system to maintain a relatively large
temperature difference across the junction. In this section, we
study the Seebeck coefficient in the linear and nonlinear regimes.
An analytical solution of $S$ in the linear response regime gives
useful guidelines for understanding the behavior of $S$ in the
nonlinear response regime.

In the linear response regime, Eq. (5) can be rewritten as
$J=L_0e\Delta V_a+ L_1k_B\Delta T$. The thermoelectric coefficient
$L_n$ is given by
\begin{equation}
L_n=\frac{2e}{h}\int d\epsilon {\cal
T}(\epsilon)\frac{(\epsilon-E_F)^n}{4cosh^2((\epsilon-E_F)/(2k_BT))},
\end{equation}
where the transmission coefficient ${\cal T}(\epsilon)$ is
calculated under the equilibrium condition. We define $S(T)=\Delta
V_a/\Delta T=(k_B/e)(-L_1/L_0)$. Figure 4 shows the Seebeck
coefficient ($S$) of SCQDs with identical QDs as a function of
temperature at various QD energy levels $E_0-E_F=\Delta=20, 30, 40,
50~\Gamma_0$. The negative $S$ indicates that the diffusion
electrons pass through the resonant channels of $E_0\pm t_c$,
$E_0+U_{12}\pm t_c$, $E_0+U+U_{12}\pm t_c$, and $E_0+U+2U_{12}$.
These resonant channels result from the four configurations of
$p_1$, $p_3$, $p_6$, and $p_8$ in Eq. (7). The maximum $S(T_0)$
increases with increasing $\Delta$. $T_0$ denotes the temperature at
which $S$ has a maximum value. In addition, $T_0$ is shifted toward
to the high-temperature regime. The dashed lines calculated using
$S_0(T)=-\Delta/T$ are employed to fit these solid lines. These
fitting curves show good agreement with the solid lines in the high
-temperature regime. The characteristic of $S_0=-\Delta/T$, which is
independent of the tunneling rates, was also determined in the case
of a single QD junction.$^{20)}$ In the appendix, we derive the
formula of $S_0(T)=-\Delta/T$ for the noninteraction case when the
conditions of $t_c/k_BT \ll 1$ and $\Gamma/k_BT \ll 1$ are
satisfied. Such results imply that the behavior of $S_0(T)$ is not
sensitive to electron Coulomb interactions. This is attributed to
the very small contribution of resonant channels involving electron
Coulomb interactions that are far above $E_F$. Note that the
behavior of $S(T)$ becomes complicated if when $E_{\ell}-E_F=\Delta
< 0$, because $S(T)$ will involve many physical parameters such as
electron Coulomb interactions, the electron interdot hopping
strength $t_c$, and tunneling rate.$^{17)}$

To calculate the Seebeck coefficient in the nonlinear response
regime, we numerically solve Eq. (5) by considering the condition of
$J=0$. Figure 5 shows the electrochemical potential and Seebeck
coefficient as a function of temperature difference $\Delta T$ for
different detuning energies $\Delta $ at $k_BT_L=40~\Gamma_0$. The
curve with triangles(considering $\Delta=50~\Gamma_0$) neglects the
shift of the QD energy level yielded by the electrochemical
potential. That is, $\eta_1=\eta_2=0$. $\Delta \mu$ increases with
increasing $\Delta T$. When $\Delta T
> 0$, $\Delta \mu=e\Delta V_a$ is negative. On the other hand, $\Delta
\mu$ is positive when $\Delta T < 0$. By Comparing the blue solid
line with the curve with triangles in Fig. 5(a), we find that under
the forward temperature bias ($T_L
> T_R$), the case of $\eta_1=\eta_2=0$ needs a larger electrochemical potential ($\Delta \mu$) to balance diffusion
carrier flow mainly through the resonant channels of $E_0\pm t_c$
from the left electrode to the right electrode. This is because the
resonant channels are always kept in the case of $\eta_1=\eta_2=0$.
Once $\eta_{\ell} \neq 0$, the electrochemical potential will
generate the off-resonant channels and the carrier diffusion flow
will be blocked. Under the reverse temperature bias ($k_BT_R >
k_BT_L=40\Gamma_0$), the difference between the blue solid line and
the curve with triangles is small even though $\Delta
T=-20\Gamma_0$. This indicates that diffusion carrier flow does
increase too much with increasing temperature bias. Figure 5(b)
shows the Seebeck coefficient corresponding to the electrochemical
potentials shown in Fig. 5(a). The behavior shown in Fig. 5(b) can
be roughly described by $S=S_0/(1-0.5\Delta T/T)$. If $\Delta T/2T
\ll 1$, we have $T_R=T_L-\Delta \mu/S_0$. This result may be useful
for the application of temperature detectors.$^{22)}$

According to the definition of $S=(k_B/e)(-L_1/L_0)$, $S$ is related
to the electrical conductance $G_e=eL_0$. We further examined the
relationship between these thermoelectric response functions. Figure
6 shows the electrical conductance ($G_e$) of SCQDs with identical
QDs ($E_{\ell}=E_0=E_F+30\Gamma_0-eV_g$) as a function of gate
voltage for different values of EPIs ($\Omega_{\ell}$) at a low
temperature $k_BT=1\Gamma_0.$ Note that we have $T_p=T$ in the
linear response. The coupling strength of EPIs can be tuned by the
phonon cavity, which modulates the phonon density of states to
change $\lambda_{\ell}=\Omega_{\ell}/\omega_0$.$^{9-11)}$ In the
absence of $\lambda_{\ell}=0$, there are eight resonant channels of
$E_0\pm t_c$, $E_0+U_{12}\pm t_c$, $E_0+U+U_{12}\pm t_c$, and
$E_0+U+2U_{12}\pm t_c$, which are labeled from $eV_{g1}$ to
$eV_{g8}$. The first peak ($eV_{g1}$) is shifted toward a low gate
voltage with increasing $\lambda$, which is attributed to the shift
of the QD energy level $E^e_{\ell}=E_{\ell}-\lambda^2 \omega_0$.
Because of the reduction of the interdot Coulomb interaction
$U^e_{\ell,j}=U_{\ell,j}-2\lambda_1\lambda_2 \omega_0$, there are
only six peaks and four peaks in the $\lambda_1=\lambda_2=0.5$ and
$\lambda_1=\lambda_2=0.7$. For $\lambda_1=\lambda_2=0.7$,
$U^e_{\ell,j}$ almost vanishes. Therefore, four peaks correspond to
$E^e_{\ell}\pm t_c$, and $E^e_{\ell}+U^e_{\ell}\pm t_c$. For the
homogenous coupling of EPIs $\lambda_1=\lambda_2$, $t^e_c=t_c$ is
independent of $\lambda$. The separation between the peaks
corresponding to the bonding and antibonding states is invariant.

In Fig. 6, we considered the case of $\lambda_1 > 0$ and $\lambda_2
> 0$. It is possible to have a positive $\lambda_1$ and a negative
$\lambda_2$ by considering QDs at particular locations in the phonon
cavity. For instance, dot A is out of phase from dot B under phonon
fields [see Eq. (4)]. This will lead to a reduction of intradot
Coulomb interaction and an enhancement of interdot electron Coulomb
interaction. Figure 7 shows the electrical conductance ($G_e$) and
Seebeck coefficient ($S_0$) as a function of gate voltage for
different $\lambda$ values. For $\lambda_1=0.5$ and
$\lambda_2=-0.5$, the spectra of $G_e$ and $S_0$ are significantly
changed owing to the reduction of interdot hopping strength [$t^2_c
e^{-(\lambda_{1}-\lambda_2)^2 coth^2(\omega_0/(2k_BT))}]$ and the
enhancement of interdot Coulomb interactions $U_{\ell,j}-2\lambda_1
\lambda_2 \omega_0$. The Seebeck coefficients shown in Figs. 7(c)
and 7(d) correspond to Figs. 7(a) and 7(b), respectively. The
Seebeck coefficient shows a change in sign arising from the bipolar
effect which is electron-hole asymmetrical, where holes are defined
as the empty states below the Fermi energy of electrodes.$^{17)}$

\textbf{4. Summary and Conclusions}

Intradot Coulomb interactions are more important than interdot
Coulomb interactions in the observation of current rectification
spectra in the Pauli spin blockade. The current rectification is not
restricted in the weak interdot hopping limit ($t_c/\Gamma \ll 1)$.
From the experimental point of view, the release from the weak
interdot hopping restriction is very useful for measuring the
tunneling current of SCQDs. Unlike the charge current rectification,
the maximum spin-polarization current is obtained in the forward
bias regime. The Seebeck coefficients in the linear and nonlinear
response cases were studied. The universal behavior of the Seebeck
coefficient ($S_0=-k_B\Delta/eT$) may be useful for the application
of temperature detectors. EPIs provide an extra degree of freedom to
control carrier transportation.

\mbox{}\\
{\bf Acknowledgments}- This work was supported by  National Science
Council, Taiwan, under Contract Nos. NSC 101-2112-M- 008-014-MY2 and
NSC 101-2112-M-001-024-MY3.

\mbox{}\\
${}^{\dagger}$ E-mail address: mtkuo@ee.ncu.edu.tw\\



\appendix
\section{Seebeck coefficient}

The tunneling current given by Eq. (5) can be analytically
calculated by contour integration. For simplicity, this appendix
only considers the case without electron Coulomb interactions.

\begin{equation}
J=\frac{2e\Gamma_L \Gamma_R}{h} \int d\epsilon
\frac{t^2_c[f_L(\epsilon)-f_R(\epsilon)]}{|(\epsilon-\epsilon_1+i\Gamma_{L}/2)
(\epsilon-\epsilon_2+i\Gamma_{R}/2)-t^2_c|^2},
\end{equation}
$f_{L(R)}(\epsilon)$ is the Fermi distribution function of the left
(right) electrode. $\epsilon_{\ell}=E_{\ell}+\eta_{\ell}e\Delta
V_a$. In the linear response regime, we take $e\Delta V_a
\rightarrow 0 $ and $k_B\Delta T \rightarrow 0$ in Eq. (A1) and we
have $J=L_0e\Delta V_a+L_1k_B\Delta T$, where

\begin{equation}
L_n/\alpha =\int_{-\infty}^{\infty} dy
\frac{y^n/cos^2(y/2)}{(y-\omega_{+})(y-\omega^*_{+})(y-\omega_{-})(y-\omega^*_{-})},
\end{equation}
where $\alpha=\frac{2e\Gamma_0^2t^2_c}{4h(k_BT)^4}$,
$\omega_{\pm}=(E_{\pm}-E_F)/(k_BT)$, and
$E_{\pm}=\epsilon_0+i\Gamma/2\pm E_{\delta}$. We define
$\epsilon_0=(E_1+E_2)/2=E_0$,
$\Gamma=(\Gamma_L+\Gamma_R)/2=\Gamma_0$, and
$E_{\delta}=\sqrt{((E_1-E_2)/2)^2+t_{c}^2}$. The Seebeck coefficient
is defined as $S_0(T)=-L_1/L_0 (k_B/e)$ for considering the
condition of $J=0$.

By using contour integration, $L_0$ and $L_1$ can be evaluated in
terms of polygamma functions.$^{23)}$ Because we are interested in
the case of $k_BT/\Gamma_0 \gg 1$ and $\Delta=(E_0-E_F) \gg t_c$, we
have the thermoelectric response functions $L_0$ and $L_1$
\begin{equation}
L_0=\frac{2e}{h} \frac{t^2_c}{k_BT} \frac{\pi
\Gamma_0/2}{t^2_c+(\Gamma_0/2)^2} \frac{1}{cosh^2(\Delta/2k_BT)}
\end{equation}
and

\begin{equation}
L_1=\frac{2e}{h} \frac{t^2_c}{k_BT^2} \frac{\pi
\Gamma_0/2}{t^2_c+(\Gamma_0/2)^2}
\frac{\Delta}{cosh^2(\Delta/2k_BT)}.
\end{equation}

On the basis of Eqs. (A3) and (A4), we obtain the electrical
conductance ($G_e $) and Seebeck coefficient ($S_0$) with the
following expressions
\begin{equation}
G_e=\frac{2e^2}{h} \frac{t^2_c}{k_BT} \frac{\pi
\Gamma_0/2}{t^2_c+(\Gamma_0/2)^2} \frac{1}{cosh^2(\Delta/2k_BT)}
\end{equation}

and

\begin{equation}
S_0=\frac{-k_B\Delta}{eT}.
\end{equation}
The Seebeck coefficient of Eq. (A6) depends on the detuning energy
and equilibrium temperature.

\newpage

{\bf Figures and Figure Captions}
\begin{figure}[h]
\centering
\includegraphics[scale=0.3]{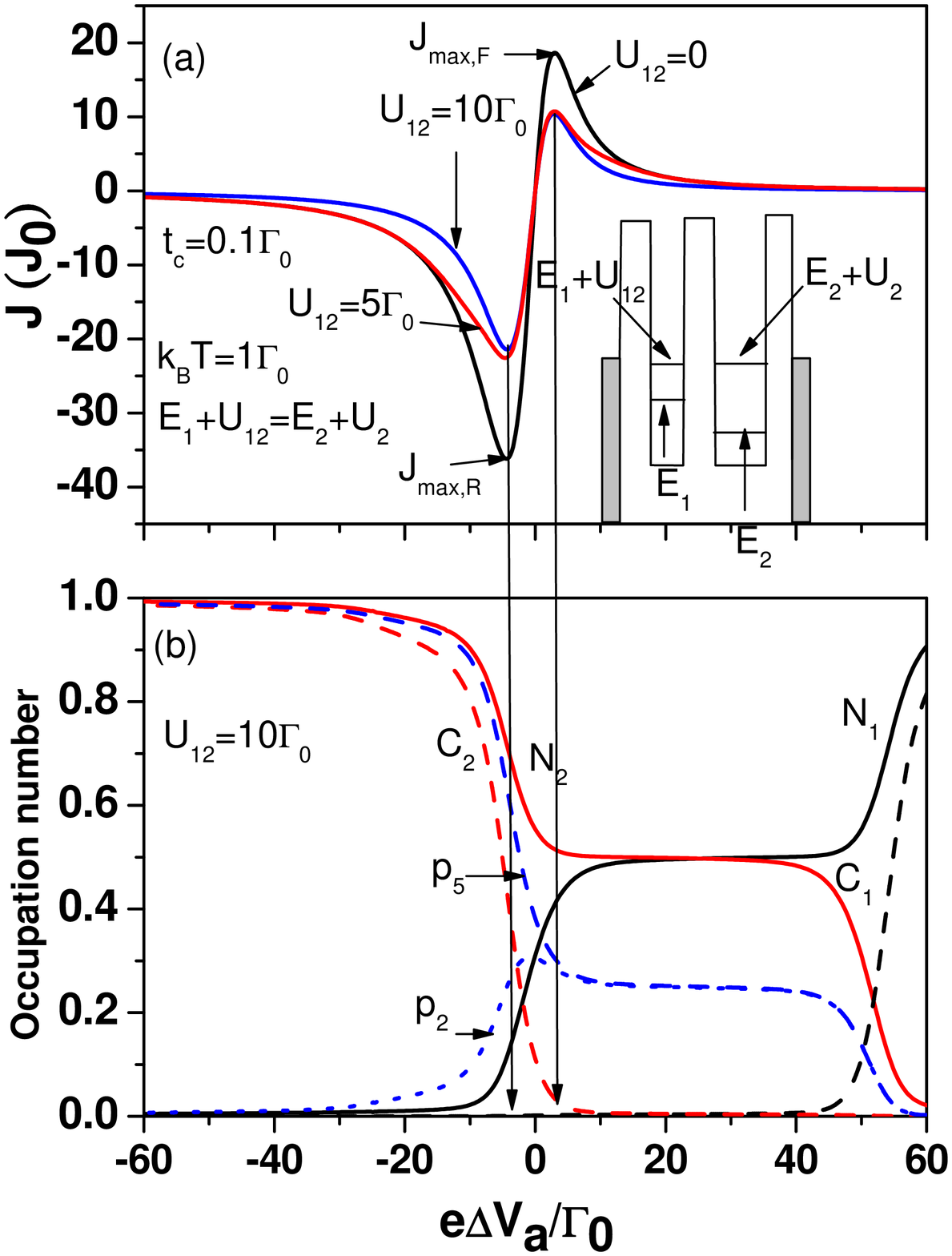}
\caption{(a) Tunneling current as a function of applied bias for
various interdot Coulomb interactions ($U_{12}$) at temperature
$k_BT=1\Gamma_0$. The tunneling currents are in the units of
$J_0=2e\Gamma_0/h$. (b) Occupation number ($N_{\ell}$) and two
-particle correlation function ($c_{\ell}$) in the case of
$U_{12}=10\Gamma_0$.}
\end{figure}

\begin{figure}[h]
\centering
\includegraphics[scale=0.3]{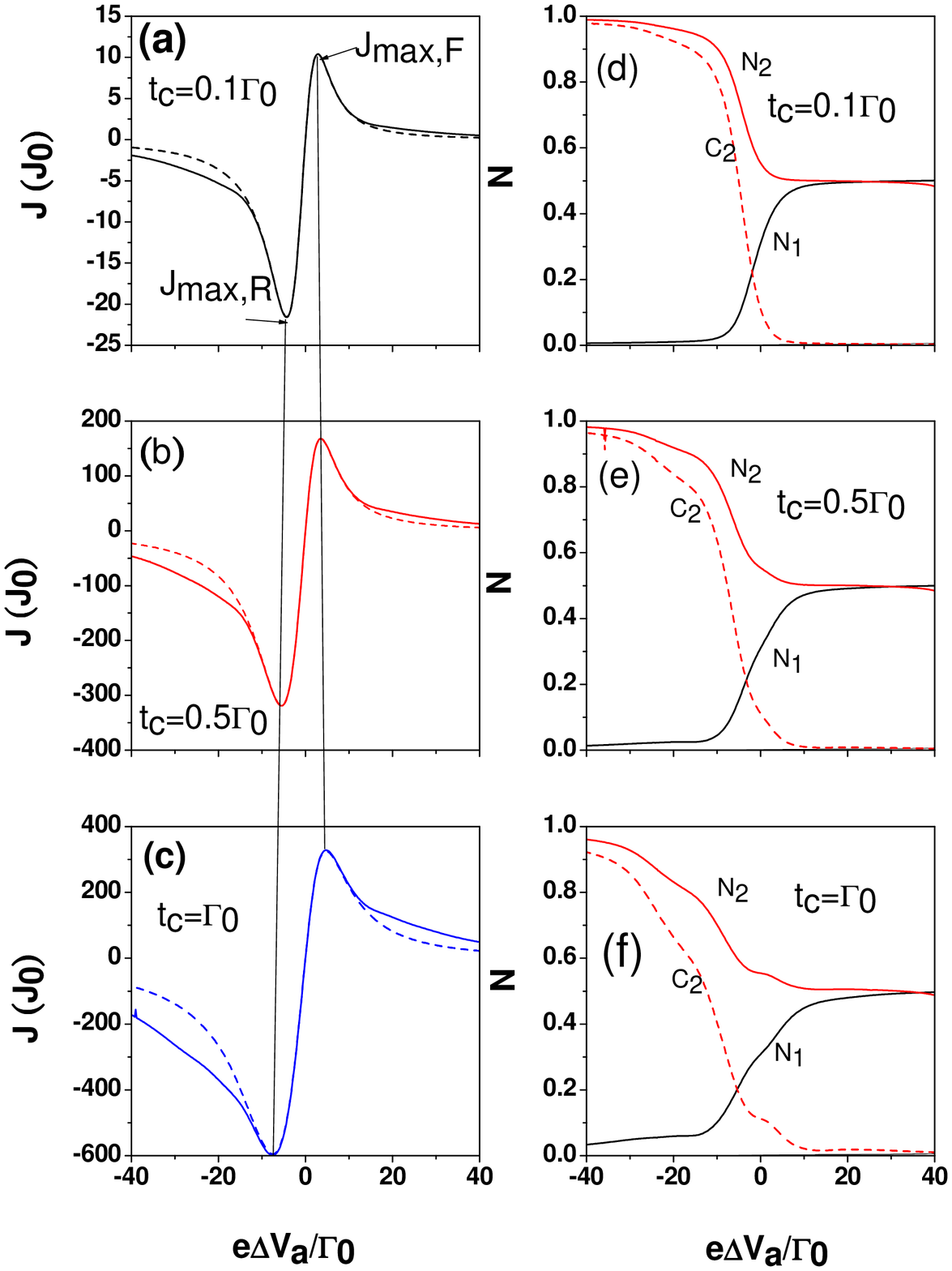}
\caption{Tunneling current as a function of applied bias. Diagrams
(a)-(c) show the interdot hopping strengths $t_c=0.1,0.5$, and
$1~\Gamma_0$, respectively. Diagrams (d)-(f) showing $N$ and
$c_{\ell}$ correspond to diagrams (a)-(c), respectively. Note that
$c_1$ is not plotted in diagrams (d)-(f) because of the very small
$c_1$.}
\end{figure}

\begin{figure}[h]
\centering
\includegraphics[scale=0.3]{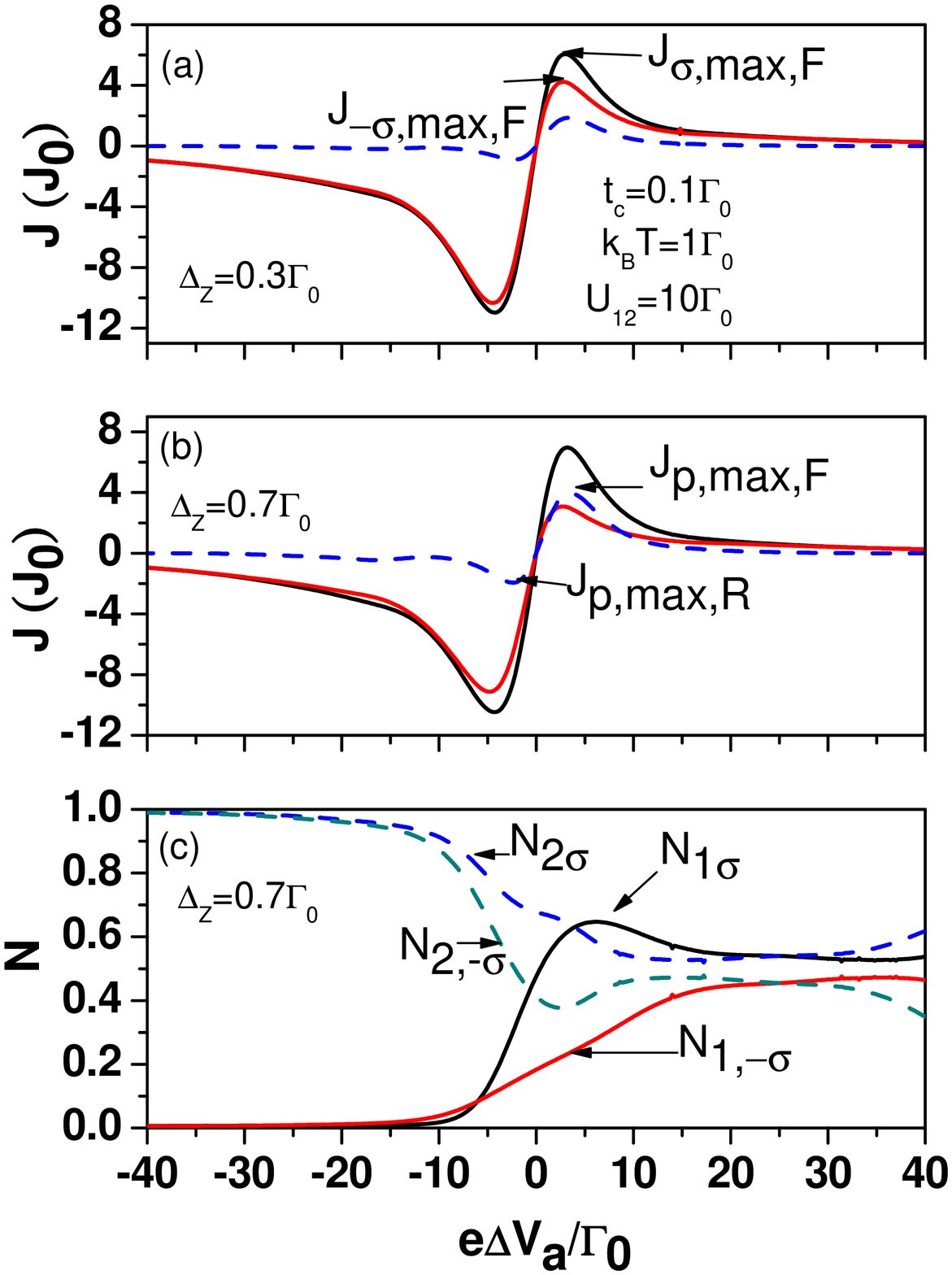}
\caption{Spin-dependent tunneling current as a function of applied
bias: (a) $\Delta_Z=0.3~\Gamma_0$ and (b) $\Delta_Z=0.7~\Gamma_0$.
$\Delta_Z=g\mu_BB/2$. The spin-polarization currents (dashed lines)
are defined as $J_p=J_{\sigma}-J_{-\sigma}$. (c) Spin-dependent
occupation number as a function of applied bias at
$\Delta_Z=0.7\Gamma_0$.}
\end{figure}

\begin{figure}[h]
\centering
\includegraphics[scale=0.3]{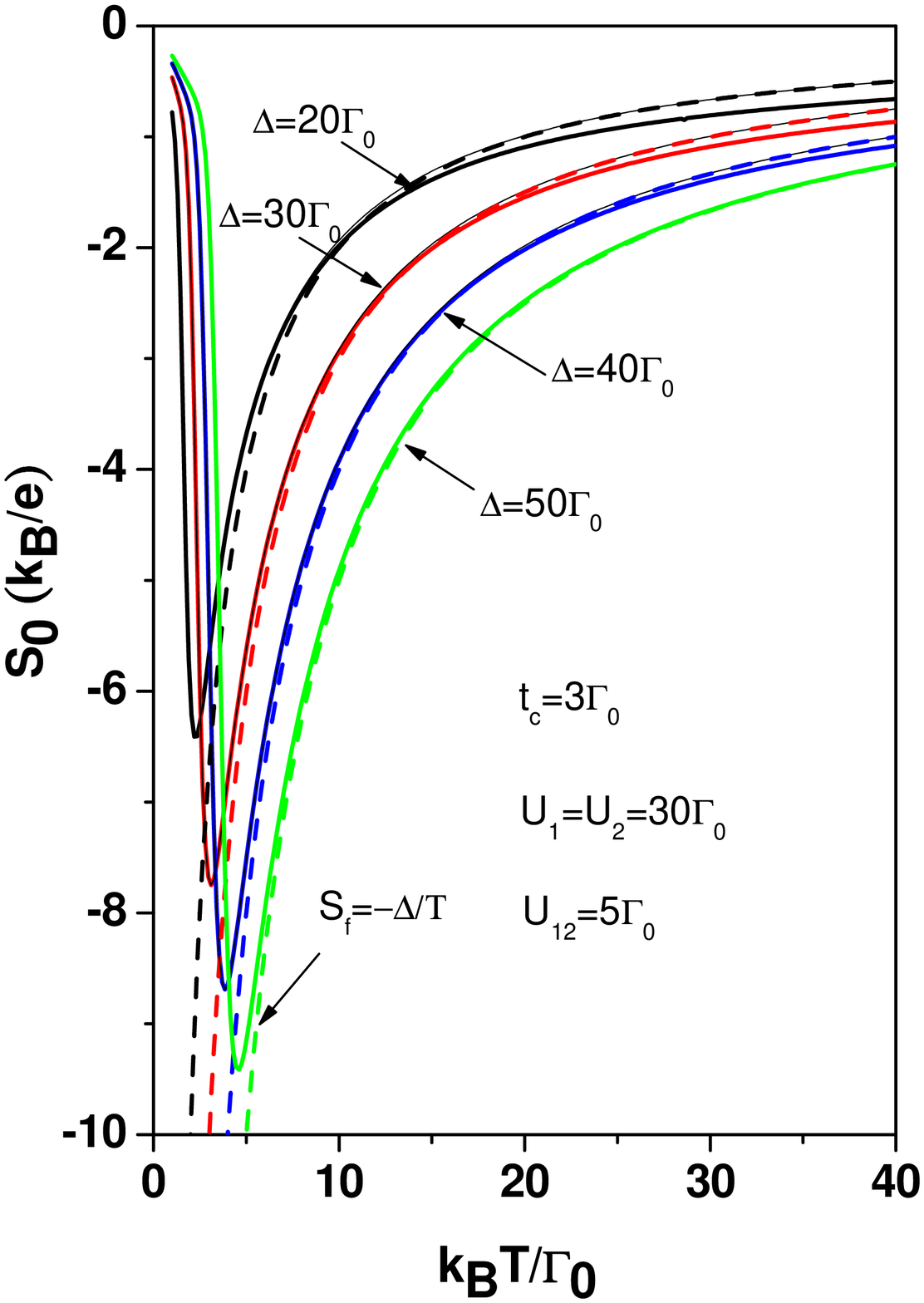}
\caption{Seebeck coefficient ($S_0$) as a function of temperature
for different detuning energies $(\Delta=20,30,40,50 ~\Gamma_0$) in
the linear response regime. The physical parameters
$U_{\ell}=U=30\Gamma_0$, $U_{\ell,j}=5\Gamma_0$,
$\Gamma_L=\Gamma_R=\Gamma_0$, and $t_c=3\Gamma_0$ are adopted.
Dashed lines are calculated using $S_f=-\Delta/T$, which is given by
Eq. (A6).}
\end{figure}

\begin{figure}[h]
\centering
\includegraphics[scale=0.3]{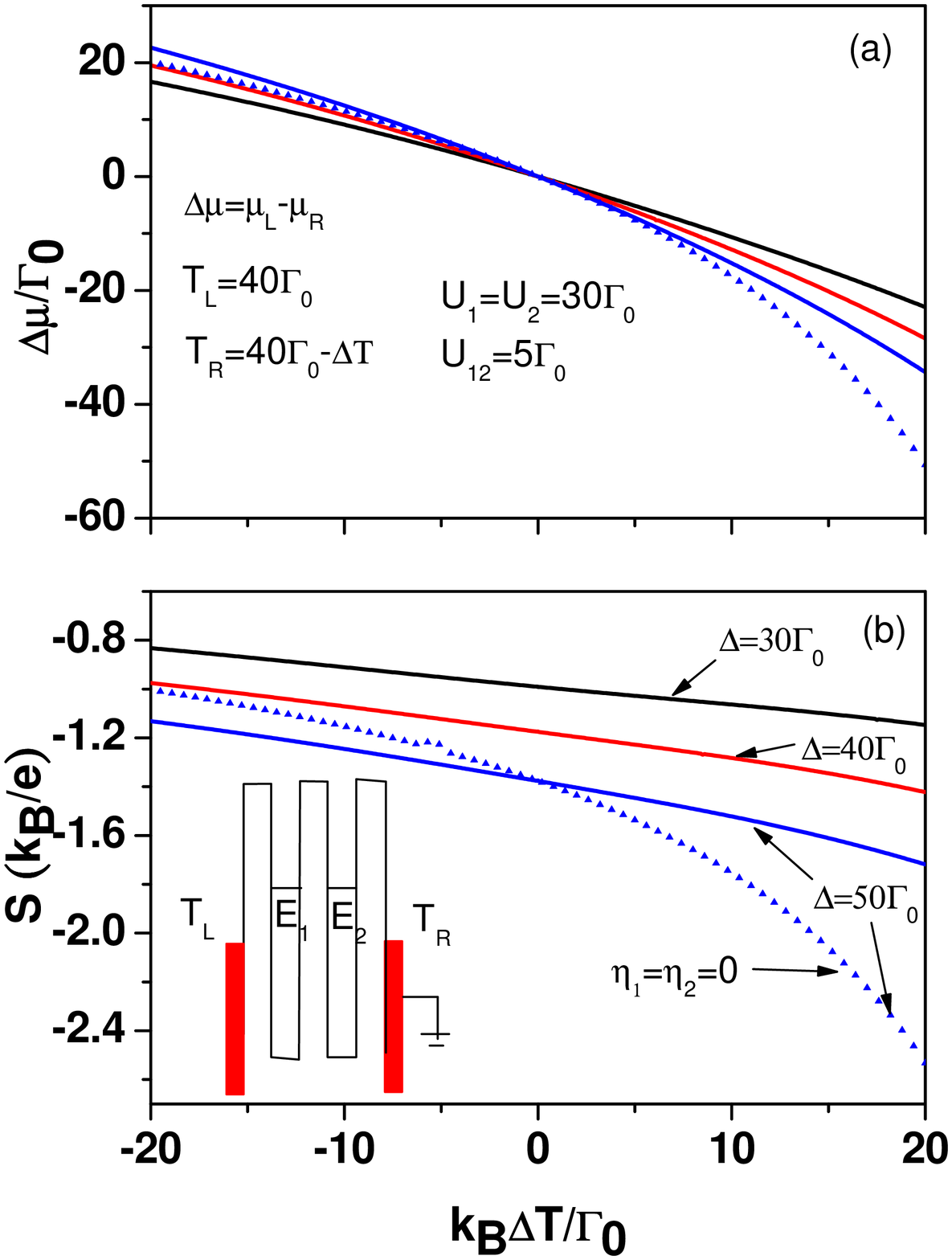}
\caption{(a) Electrochemical potential $\Delta\mu$ and (b) Seebeck
coefficient (S) as a function of temperature difference $\Delta T$
for different detuning energies at temperature of the left electrode
$T_L=40\Gamma_0$. Other physical parameters are the same as those in
Fig. 4.}
\end{figure}

\begin{figure}[h]
\centering
\includegraphics[scale=0.3]{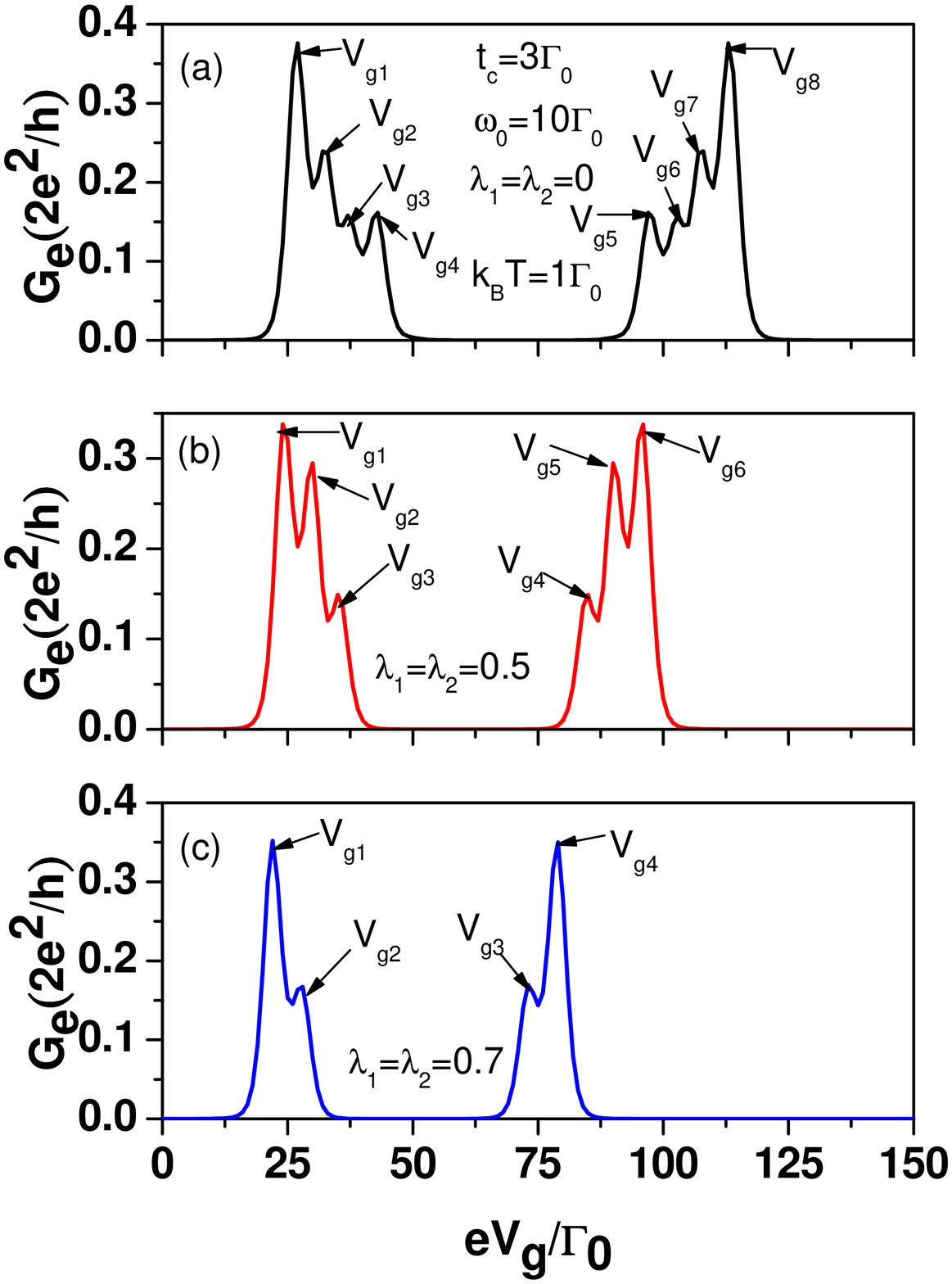}
\caption{Electrical conductance ($G_e$) as a function of applied
gate voltage for different electron-phonon interactions (EPIs) at
$k_BT=1\Gamma_0$, $E_{\ell}=30\Gamma_0-eV_g$, $U_{\ell}=60\Gamma_0$,
and $U_{\ell,j}=10\Gamma_0$. Diagrams (a)-(c) show the results for
$\lambda_1=\lambda_2$=0, 0.5, and 0.7, respectively}
\end{figure}

\begin{figure}[h]
\centering
\includegraphics[scale=0.3]{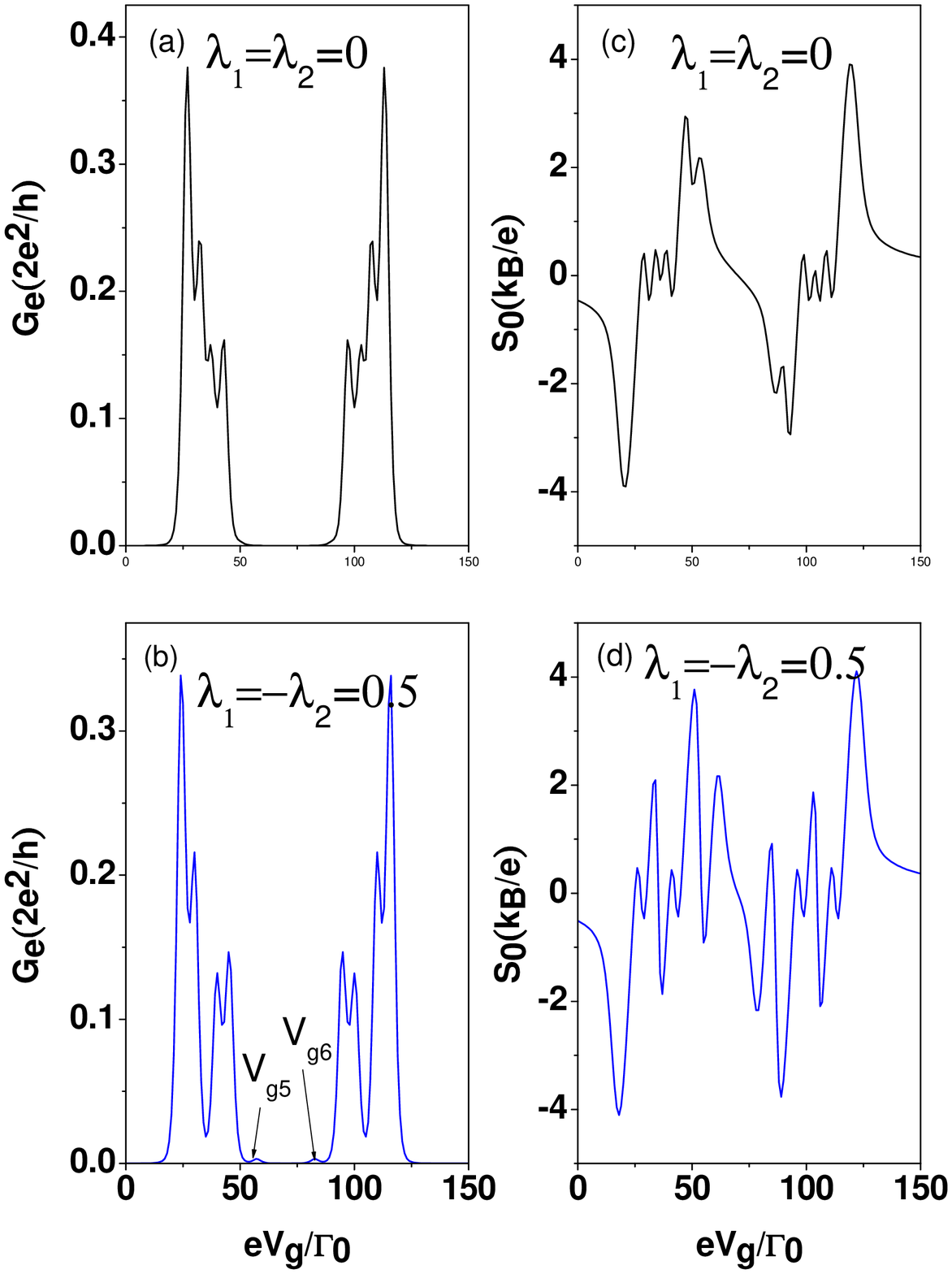}
\caption{(a)-(b) Electrical conductance ($G_e$) and (c)-(d) Seebeck
coefficient $(S_0)$ as a function of applied gate voltage for
different EPIs at $k_BT=1\Gamma_0$. Solid lines:
$\lambda_1=\lambda_2=0$, dashed lines: $\lambda_1=-\lambda_2=0.5$.
Other physical parameters are the same as those in Fig. 6.}
\end{figure}



\end{document}